# Synthesis, processing and transport properties of cubic and trigonal solid solutions in the Mg-Si-Sn system

Pathikumar Sellappan[1] Anthony Y. Fong,[1,#] Masayuki Murata,[2] Yasuhiro Kodera[1] and Javier E. Garay[1*]

[1] Advanced Materials Processing and Synthesis (AMPS) Lab, Materials Science and Engineering Program and Department of Mechanical & Aerospace Engineering, University of California San Diego, La Jolla, CA 92093 USA

[2] Thermoelectric Energy Conversion Group, National Institute of Advanced Industrial Science and Technology (AIST), Tsukuba, Ibaraki 305-8586, Japan

\* - Corresponding author
\# - currently at Stanford Synchrotron Radiation Lightsource, Menlo Park, CA

## Abstract

Mg-Si-Sn solid solutions have interesting and useful semiconducting properties particularly for thermoelectric applications. While cubic solid solutions have been prepared by a variety of methods, solid solutions with a trigonal crystal structure have not been reported. We employed a combination of high energy ball milling and current activated pressure assisted densification (CAPAD) to induce and complete phase transformation from cubic to trigonal phase, forming dense trigonal composites. This is the first successful preparation of the trigonal phase in Mg-Si-Sn stem. In addition, we report the transport properties of both the cubic and trigonal composites measured in the 160 to 400 K range. The electrical conductivity of the trigonal composites was several times higher compared to the cubic counterparts while thermal conductivity is lower but in the range of previously reported Mg-Si-Sn materials. The Seebeck coefficient of the trigonal composites is lower compared to the cubic samples, leading to low *ZT* values. The *ZT* of the cubic samples, however is higher than previously reported for un-doped Mg-Si-Sn solid solutions which is promising for thermoelectric applications.

## 1. Introduction

Magnesium based Zintl compounds with composition $Mg_2X$ where *X* is for example Sn, Si have been studied for decades because of their interesting and useful semiconducting properties.[1–3] These compounds have a cubic antifluorite structure under ambient pressures and temperatures and they readily form ternary solid solutions by substitution on the *X* site.[4–6] Ternary solid solutions (Mg-Si-Sn) are promising thermoelectric materials since solid-solution formation can significantly reduce thermal conductivity due to the mass difference between Sn and Si and hence increase thermoelectric performance.[7–9] Particularly enticing is that the Mg, Sn and Si are earth-abundant, inexpensive and can be obtained in relatively earth friendly methods. In addition, they exhibit high power output per unit mass compared to many other TE materials due to their low density values.[10–12] These attributes have caused intense interest in $Mg_2Si$-$Mg_2Sn$ solid solutions as thermoelectric (TE) materials for large scale commercial applications.[1,12–14]

The ambient temperature phase diagrams of these Zintl phases suggests that the $Mg_2Sn$ and $Mg_2Si$ are the only stable compound (only compound present) and that they are line compounds.[4] A denser, trigonal 'high pressure' binary phase in the Mg-Sn has been reported, but it has been understudied because of its processing difficulties-high pressures (> 1 GPA) are typically required for synthesis.[15,16] This phase is often reported to be slightly Mg deficient compared to cubic $Mg_2Sn$ having a $Mg_9Sn_5$ stoichiometry. More recently, through a combination of high energy ball milling and current activated pressure assisted densification (CAPAD) we were able to produce dense polycrystalline samples composed primarily of the trigonal phase in the Mg-Sn system.[17] In contrast to the Mg-Sn system, a trigonal phase has not been reported in the Mg-Si system or in Mg-Si-Sn solid solutions. Since solid solution formation is possible in the cubic system we surmised that trigonal solid solutions might also exist. Here we report a successful synthesis and processing procedure for cubic and trigonal composites in the Mg-Si-Sn ternary system and report their transport properties (electrical conductivity, Seebeck coefficient, thermal conductivity and thermoelectric figure of merit, *ZT*). We believe this is the first report of the trigonal phase in the ternary Mg-Si-Sn system.

## 2. Experimental procedure

### 2.1. Powder synthesis and consolidation

Mg-Si-Sn alloys powders with a composition of $Mg_2Si_{0.1}Sn_{0.9}$ were synthesized by solid-state reaction method. The composition was chosen to be on the Sn rich side of Mg-Si-Sn system. No intentional dopants were added in the current study. Elemental Mg, Si and Sn powders (99.8 % metal purity, 325 Mesh Size, Alfa Aesar) were mixed thoroughly using agate mortar and pestle, sealed in an Ar filled tube furnace, and heat treated under flowing Ar atmosphere at 600 °C for 6 h. Due to high volatile nature of Mg, additional 10 at% Mg was added to the mixture to compensate for the loss during the synthesis and processing. Heat treated (reacted) powders were then crushed, sieved through 325 mesh (~44 um). Some of the as synthesized powder was then planetary ball milled (Fritsch Pulverisette 7) at 450 RPM using stainless steel jars and balls for varying time. The ball to powder weight ratio was 10:1. All powders and milling media were handled in glove-box. After the ball milling, the powders were crushed and sieved again for consolidation and characterization.

Both the high-energy ball milled as well as as-synthesized (without the high-energy ball milling) powders were consolidated into dense specimens using CAPAD (also referred to as Spark Plasma Sintering (SPS)).[18] Detailed description on the experimental conditions are outlined elsewhere.[17] A custom built CAPAD setup was employed to produce 9.52 mm diameter cylindrical samples approximately 2.5 mm thick.[18] All samples were densified under 112 MPa pressure and temperatures up to 700 °C with a heating rate of 200 °C/min under vacuum (<$10^{-2}$ Torr). The samples were held for 5 minutes at the final holding temperature for all the cases. Entire sample fabrication duration was less than 20 minutes. The temperature was monitored using a type-N thermocouple placed in a non-through hole half way into the die wall.

### 2.2 Structural Characterization

The phase content of the powders and bulk CAPAD samples was characterized using XRD analysis (PANalytical Empyrean diffractometer using a Cu Kα X-ray source ($K_{\alpha 1}$ = 1.54056 Å and $K_{\alpha 2}$ = 1.54440 Å). Bulk densities of the CAPAD processed specimens were measured using the Archimedes method. Polished as well as fractured surface of the samples were microstructurally investigated using scanning electron microscopy (SEM, Zeiss Sigma 500). No conductive coating was applied on the sample surfaces prior to the SEM analysis.

**2.3 Transport property measurement**

Selected CAPAD processed specimens were sectioned into rectangular bar specimens using low-speed diamond saw and all sides were polished using 1200 grit SiC paper to a achieve smooth surface finish. The dimensions of measured samples were $1.33 \times 2.25 \times 7.15$ mm$^3$, $1.56 \times 2.28 \times 7.69$ mm$^3$, $1.50 \times 2.53 \times 7.01$ mm$^3$ and $1.27 \times 2.22 \times 6.80$ mm$^3$ for No PBM III, PBM 1, PBM 2 and PBM 3, respectively. A custom-built setup was employed to measure the electrical resistivity, Seebeck coefficient and thermal conductivity from 160 to 400 K, which were precisely controlled by a Stirling cooler and heater. The measurement set-up in this study was configured based on the previous work.[19] The electrical current for the electrical conductivity measurement, the temperature difference of the Seebeck coefficient and thermal conductivity measurement were all applied along the longitudinal direction of the sample. The magnitude and frequency of the electrical current needs to be selected carefully to prevent Joule heating and the phase delay, respectively.[20] In this case, an electrical current of 1 mA (No PBM III), and 10 mA (PBM 1, PBM 2, PBM 3) was applied and the frequency of 11.234 Hz was selected for all samples. The Seebeck coefficient and thermal conductivity were measured by the steady-state method at the same time. The ZT values were calculated form measured results according to:

$$ZT = \frac{\sigma S^2 T}{k} \qquad \text{(Eq. 1)}$$

where ($\sigma$) is electrical conductivity, $S$ is Seebeck coefficient and $k$ is the thermal conductivity.

## 3. Results & Discussion

The high-energy planetary ball milling (PBM) studies on the ternary (Mg$_2$Si$_{0.1}$Sn$_{0.9}$) powders are shown in **Fig 1**. The XRD results shows that the as-synthesized powders (labeled as as-synthesized 0 h) was phase pure cubic. **Fig 1a** shows that the ball milling has profound effect on the phase content of the powder. The most intense cubic peak (111) at 22.9° 2θ starts to decrease with increasing milling time, demonstrating that there is a phase transformation from cubic to trigonal with increasing PBM time. This cubic to trigonal transformation is similar to what has been previously observed for the binary Mg-Sn system.[17,21] **Fig 1a**, shows that phase transformation is attained after the relatively long time of 27 h. Continuing the milling, the trigonal phase starts to be the dominate phase in the composites by 51 h. Further ball milling for 75 h results in a powder composition with majority of trigonal phase with minor cubic phase, and some minor XRD detectable elemental Sn and MgO. These results are comparable to previous reports for mechanical milling resulting in majority trigonal phase in the Mg-Sn system near the Mg$_2$Sn composition.[17] We believe that this is the first evidence of the trigonal phase in the Mg-Si-Sn system. It should be noted that the trigonal phase was not observed in the ternary system with higher concentrations of Si even after similar milling time (results not shown here).

The evolution of the cubic to trigonal transformation is more easily appreciated in **Fig 1b** that plots the integral intensity ratio of primary cubic peak (111) to the primary trigonal (300) at 23.36° 2θ. **Fig. 1a** and **1b** confirm high-energy ball milling process effectively promotes the formation of trigonal phase in the ternary Mg-Si-Sn system. The powders that were ball milled for 75 hours were then crushed, sieved and consolidated using CAPAD.

XRD patterns of the CAPAD specimens consolidated at various temperatures are shown in **Fig. 2**. The XRD of the samples consolidated from as-synthesized powder i.e. those with no PBM (**Fig. 2a**) showed no signs of phase transformations—the powder was cubic and the densified samples are all cubic as well. By contrast, XRD patterns clearly reveal that the samples consolidated at 355, 415 460 °C from PBM powder are composites consisting of both cubic and trigonal phases (**Fig. 2b**). CAPAD processing at 500 °C results in a specimen with only trigonal phase and no evidence of cubic phase can be seen. **Fig. 2c are** XRD patterns of the consolidated samples in the region of the most intense peak. XRD of previously published trigonal, Mg-Sn binary specimen[17] is plotted for comparison. There are small peak shifts towards higher angle in the ternary (Mg-Si-Sn) *vs* the binary system (Mg-Sn) suggesting that the smaller Si atom incorporation into the trigonal solid solution since the Sn atomic radius (1.40 x 10$^{-10}$ m) is larger than the radius of Si (1.176 x 10$^{-10}$ m).

The integral intensity ratio of primary trigonal peak (300) over primary cubic peak (111) shown in **Fig 2d** clearly reveals the completion of phase transformation. Also, the results show that the cubic to trigonal transformation continues during CAPAD processing. However, similar to the powder samples (**Fig. 1b**), the bulk, dense specimens also contain XRD detectable elemental Sn and MgO even though extensive care was taken during the processing and handling. The high energy ball milled, trigonal phase transformed samples are labelled as PBM 0, PBM 1, PBM 2, PBM 3. The CAPAD specimen processed form as-synthesized, no PBM powders is labelled as No-PBM I, II and III. The processing condition as well as the phase present information are provided in **Table 1**. These data confirm our hypothesis that dense, bulk trigonal solid solutions of the Mg-Si-Sn system can be produced through PBM and CAPAD.

The measured density values are plotted against the CAPAD processing temperature in **Figure 3**. The theoretical density values for the available cases are also indicated using dotted lines. Theoretical density values of the binary, cubic systems $Mg_2Sn$ (3.599 g/cm$^3$), $Mg_2Si$ (1.988 g/cm$^3$) and the trigonal-$Mg_9Sn_5$ (3.997 g/cm$^3$) are calculated using lattice parameters/volume obtained from ICSD references 00-065-2997, 00-034-0458, and 01-073-8010 respectively. Theoretical density value of the cubic ternary with the $Mg_2Si_{0.1}Sn_{0.9}$ composition is calculated using lattice parameters from the measured XRD pattern of the as-synthesized powders (600 °C/6 h). The data shows that the bulk density of the CAPAD pressed samples increase with the CAPAD processing temperature for both phase pure cubic (from powder that was not ball milled) and trigonal composites (from PBM powder).

The role of high-energy planetary milling after the solid-state synthesis is clearly seen in the density plots of the dense specimens. The density of the phase pure cubic (No PBM) samples that were produced from powder with no ball milling, CAPAD processed at 600 °C was 90% and increased to ~92% when the densification temperature was raised to 700 °C. On the other hand, the trigonal composites can be densified at much lower temperatures (< 500 °C) which is likely due to the high energy ball milling step. Mechanical activation by ball milling has been observed previously to significantly decrease the temperature required for densification on other materials system.[22,23] Also, the ternary trigonal composites have significantly higher bulk densities than their cubic counter parts which is consistent with previous reports on binary Mg-Sn,[17] showing that the trigonal density is higher than the cubic.

The relative density values of trigonal composites reaches near theoretical density values when the CAPAD processing condition reaches 460 °C. Since there is no previously reported trigonal phase in the Mg-Si-Sn ternary system, there is no reference available to compare density values. However, adding Si to the trigonal solid solution, *i.e*, making trigonal Mg-Si-Sn should result in a decrease in mass density compared to the binary Mg-Sn. This is consistent with the difference in the density values of Mg-Si-Sn and Mg-Sn cubic system (**Fig. 3**). Hence, we assume that the theoretical density values of the trigonal $Mg_2Si_{0.1}Sn_{0.9}$ composition lies just below the theoretical density value of trigonal Mg-Sn. The measured bulk density values of trigonal composites peaks at 460 °C and decreases slightly at 500 °C. We observed evidence of a small amount of material melting present outside of the die at this temperature. Thus, we believe that this density decrease is due to localized melting of small amounts of elemental Sn present in the system which results removal of the Sn which is the highest density constituent among the three elements. Hence, the sample processed at 500 °C (PBM 3) is likely slightly off from the chemical composition compared to the other specimens (415 and 460 °C).

Microstructure analysis of polished cross-sections of CAPAD processed dense specimens are shown in **Figure 4**. The secondary electron (SE) images and the back scattered images (BSE) are taken at the same location. Micrographs of the cubic $Mg_2Si_{0.1}Sn_{0.9}$ consolidated at 700 °C (No PBM III) are shown in **Fig 4a and b**. Since the XRD of this sample (**Fig. 2**) shows only cubic peaks we assume that the contrast is due to lack of matter (*i.e,* porosity) and not phase differences. Thus, the dark spots shown in the micrograph are pores which is consistent with the density measurements that show about 10% porosity. The average pore size is 8.2 µm (range: 4.1 to 22.5 µm). **Fig. 4c** and **4d** are micrographs of PBM 1 - processed at 415 °C. Comparison of this trigonal + cubic composite to the pure cubic sample shows a significant decrease in porosity, which is again consistent with the density measurements (**Fig. 3**). PBM 1 has an average pore size of 2.8 µm (range: 1.3 to 6.65 µm). The trigonal + trace amount of cubic sample (**Fig. 4 e** and **f**) which was processed at 460 °C (PBM 2) also appears dense with an average pore size of 0.98 µm (range: 0.52 to 1.8 µm). The trigonal sample processed at 500 °C (PBM 3) has the smallest average pore size of 0.44 µm and pores range from 0.14 to 1.0 µm (**Fig. 4g, h**). Fracture surfaces of the all trigonal composites reveal very fine features and suggests that the grain size should be in the order of ~ 50 nm. By contrast, the cubic samples we estimate that the average grain size is much larger (~ 2- 5 µm). Micrographs of the No PBM III (700 °C) and PBM 3 (500 °C) is shown in **Figure 5a** and **5b** respectively.

**Figure 6** shows the temperature dependence of the electrical conductivity, $\sigma$, measured from 160 to 400 K of selected no-PBM (cubic) and PBM composites. All samples display an increase in $\sigma$ with increasing *T*, indicating semiconductor behavior. In general, the trigonal nanocomposites show much higher $\sigma$ (**Fig. 6a**). In addition, among the trigonal composites there is a significant increase in electrical conductivity as the relative amount of trigonal phase fraction increases. Since the MgO present in all the studied specimen is nearly identical, we believe that the changes in the transport properties of the specimens are primarily caused differences in cubic to trigonal content of the composite samples.

In order to compare the electrical conductivity of these samples to previous work, we plotted the room temperature values (~ 300 K) along with previously reported work across a wide compositional range (**Fig. 6b**). The electrical conductivity values of the phase pure cubic material (No PBM III) falls well within the range of conductivity of previously reported cubic materials.[24–26] However, the trigonal composites values show a several fold increase in conductivity compared to previous reports. It should be emphasized that all previous measurements were on the cubic phase, thus the conductivity of the trigonal composites is significantly higher compared to their cubic counterparts.

The temperature dependence of $\sigma$ in semiconductors is typically described by an Arrhenius relation in the form of:

$$\sigma = \sigma_o \exp\left(\frac{-E}{k_B T}\right) \qquad (Eq.\ 2)$$

where $\sigma_o$ is a pre-exponential constant, $k_B$ is Boltzmann's constant and $E$ is an activation energy ($E = E_g/2$ where $E_g$ is the band gap for intrinsic behavior, $E = E_a$ is the acceptor level for p-type and $E = E_g - E_d$ where $E_d$ is the donor level for n-type conductors). An Arrhenius style plot for measured conductivities is shown in **Fig. 6c and d**. The data show a linear behavior at low temperatures and plateau at high $T$ for the cubic sample (No PBM III). The low temperature linear portion yield an $E = 0.02$ eV which is significantly lower than the 0 K band gaps of both Mg$_2$Sn ($E_g = 0.35$) and Mg$_2$Si ($E_g = 0.77$).[26] Assuming n-type behavior (according to sign of $S$, discussed below) the plateau corresponds to the so called exhaustion range and the low temperature extrinsic behavior yields $E_g - E_d = 0.02$ eV. Even though the samples are not intentionally doped, extrinsic behavior is not surprising given that the starting powders are only 99.8% pure and also from milling media. By contrast, the $T$ dependence of the trigonal samples show no clear plateau (exhaustion or saturation behavior) in the measured range and the $E$ extracted from linear fits over the range is significantly lower; average values are $E = 0.007, 0.005$ and $0.002$ eV for PBM 1, 2 3 respectively. These energy levels are also likely to be extrinsic and if we assume p-type conduction represent acceptor levels, $E_a$.

**Figure 7** shows the temperature dependence of the Seebeck coefficient (S) of trigonal composites as well as cubic Mg$_2$Si$_{0.1}$Sn$_{0.9}$. **Fig. 7a** shows that the Seebeck coefficient of the cubic sample (No PBM III) is negative for the entire measured temperature range. The values decrease with increasing temperatures and reach a plateau of ~375 µVK$^{-1}$ at about room temperature. As shown in **Fig. 7a**, the Seebeck coefficient of the trigonal composites are very different; They are all positive and they have a monotonically increasing trend with temperature. This change in sign of the Seebeck coefficient values for cubic and trigonal composites suggest that primary carrier changes from electrons to holes (n-type to p-type conductor).

The trigonal composites all have very similar Seebeck values, causing them to overlap on the scale of **Fig. 7a**, however, their minor differences can be seen in **Fig. 7b** where only the PBM samples are plotted. Among the trigonal composites, the sample with more cubic composition (PBM 1) has higher Seebeck Coefficient at 160 K (15.42 µVK$^{-1}$) but a lower value at 400 K (34.11 µVK$^{-1}$). In comparison, the pure trigonal sample (PBM 3) has lower value at 160 K (13.56 µVK$^{-1}$) but increases linearly with increasing temperature reaching a higher value at 400 K (37.76 µVK$^{-1}$). The samples that contain some cubic content (PBM 1 & PBM 2) display a slight saturation at higher temperatures compared to the pure trigonal (PBM3).

Similar to the Seebeck coefficients, the thermal conductivities, $k$ of the trigonal composites show an opposite trend with temperature compared to their cubic counterparts as shown in **Figure 8**. In general, the cubic sample shows a decrease in thermal conductivity while the trigonal composites show an increasing trend. The thermal conductivities of the samples are relatively low, ranging from 1.73 to 4.95 W/(m.K) at 160 K. The trigonal composites have the lowest values and the phase pure cubic sample has the overall highest value. The decreasing temperature dependence in the cubic system has been observed before in nominally un-doped cubic Mg-Si-Sn by several groups.[25–27] From kinetic theory we can write thermal conductivity,

$$k = \frac{1}{3}C\Lambda v, \qquad \text{Eq (3)}$$

where $C$ is the volumetric heat capacity, $\Lambda$ is the mean free path and $v$ is the group velocity of the heat carriers. Decreasing thermal conductivity with $T$ is indicative a decrease in $\Lambda$ caused by phonon-phonon (Umklapp) scattering (expected to have a $1/T$ dependence). By contrast, the increasing thermal conductivity trend with $T$ that we observe in trigonal composites indicates that Umklapp scattering does not dominate. This increase in $k$ with $T$ might be due to nanostructed grain size of the trigonal composites for which phonon boundary scattering should dominate.[28] Boundary scattering also reduces $k$ by reducing $\Lambda$, but unlike Umklapp scattering, boundary scattering is expected to be temperature independent. Thus an increase in $k$ would occur in regimes where the $C$ continues to rise with $T$, *i.e.* if the measurements are below the Debye temperature. We are not aware of temperature dependent $C$ measurements or calculations of the trigonal Mg-Si-Sn phase which is not surprising given the very few studies on the trigonal phase. However, the Debye temperature of both the cubic Mg-Si and Mg-Sn phases is high and is still increasing at 400 K.[6] An increase in thermal conductivity near room temperature has been observed previously in other nanostructured systems.[28–30]

Thermal conductivity values of the cubic and trigonal composites at 300 K are compared with the previously reported measurements on the Mg-Si-Sn (nominally un-doped all cubic system) in **Figure 8b**. Pervious measurements show a thermal conductivity ranging from ~2 to ~ 6 W/(m.K).[25–27] The cubic solid-solution generally show a lower value compared to the pure single phases as expected due to phonon scattering by alloying effects.[26,27,31] The range in values measured by different groups at the at same alloy composition is very likely due to microstructural differences caused by different processing routes.[25–27] All of the samples we measured fall well within the range of previous measurements on the Mg-Si-Sn system. The phase pure cubic sample of the present study (No PBM III) shows 3.89 W/(m.K) at 300 K and all the trigonal composites exhibit significantly lower values (2.73 – 2.99 W/(m.K). At this time, it is not possible to determine whether the lower $k$ values are intrinsic to the trigonal phase because of the large microstructural differences.

We used the Wiedemann-Franz Law to estimate the relative electronic/carrier ($k_C$) and lattice contribution ($k_L$) on the overall thermal conductivity ($k$). The Wiedemann-Franz is given by:

$$k_C/\sigma = LT \qquad (Eq.\ 4)$$

where $L$ is the Lorenz number. The value of $L$ depends on the material and typically ranges from $L = 2.44 \times 10^{-8}$ W$\Omega$K$^{-2}$ as originally derived for metals and is appropriate for degenerate semiconductors to $L = 1.50 \times 10^{-8}$ W$\Omega$K$^{-2}$ for nondegenerate semiconductors. Since we do not know the band structure or the precise scattering mechanisms of our samples, we used the measured value of $S$ to estimate $L$ as suggested by Snyder and coworkers.[32]

$$L = 1.5 + \exp(-\frac{|S|}{116}) \qquad (Eq.\ 5)$$

The value of $L$ we obtained using this approach ranges from ~2.2 to 2.4 $\times 10^{-8}$ W$\Omega$K$^{-2}$ for trigonal composites to ~1.6 $\times 10^{-8}$ W$\Omega$K$^{-2}$ for cubic samples. The lattice and carrier conductivity for the samples are given in **Figure 9**. The lattice thermal conductivity ($k_L$) has very similar trend to the overall thermal conductivity, $k$ (**Fig. 9a**). In addition, comparison reveals that the lattice contribution dominates the overall thermal conductivity behavior of cubic as well as trigonal composites. However, the carrier contribution $k_C$ plays a more significant role in the trigonal composites compared to the cubic sample. The data also shows that the relative carrier contribution increases with increasing trigonal content in the samples (**Fig. 9b**). The relatively high carrier contribution of the trigonal composites is consistent with the lower absolute value of the Seebeck coefficient measured in the trigonal composites compared to the cubic case (**Fig. 7**).

The thermoelectric figure of merit (*ZT*) as calculated using Eq. 1 from the measured $S$ and $k$ and $s$ for cubic as well as trigonal composites are shown in **Figure 10a**. The *ZT* values of the cubic sample increase with increasing temperature reaching a plateau at about 350 K at ~ 0.1. These values are lower than the state of the art doped Mg-Si-Sn (usually Bi and/or Sb-doped) samples that have *ZT* values near room temperature of about ~ 0.2 – 0.3.[33–36] It is possible that the *ZT* value of the cubic samples would increase with optimized doping and microstructure engineering, making them attractive for room temperature applications. For comparison, near room temperature (325 K) *ZT* values of the present samples are plotted along with previously reported, un-doped Mg-Si-Sn samples (**Fig. 10b**). The trigonal composites have comparable *ZT* values that of previously reported cubic counterparts.[25–27] Interestingly, the comparison clearly shows that our cubic Mg-Si-Sn sample has *ZT* values several times higher than previous un-doped samples. The enhanced *ZT* value of the cubic sample might be the results of its porous microstructure (**Fig. 4a**) causing relatively low thermal conductivity as recently shown by Ning et al.[37] on similar systems.

The trigonal composites *ZT* values are lower than the cubic phases and increase monotonically with *T* in the temperature range measured (**Fig. 10a**). The thermal conductivity of the trigonal composites is low and their electrical conductivity is quite high, both favorable for a high *ZT*. However as mentioned earlier, the magnitude of their Seebeck coefficients are low resulting in a lower *ZT* than the cubic Mg-Si-Sn material. The relatively high electrical conductivity, relatively high carrier thermal conductivity and low Seebeck of the trigonal composites suggest that the carrier concentration is higher than optimal. Therefore, unlike the cubic sample, it is improbable that doping would be an effective means to increase the *ZT* of the trigonal composites.

## 4. Summary

In summary, we used planetary ball milling and CAPAD to produce dense ternary solid solutions with the $Mg_2Si_{0.1}Sn_{0.9}$. The samples consolidated from powder that was not ball milled were all cubic phased, while those consolidated form PBM powder contained cubic and trigonal phases i.e. are composites. This is the first report of the trigonal phase in the Mg-Si-Sn system. Optimized conditions yields dense, bulk, trigonal $Mg_2Si_{0.1}Sn_{0.9}$. Transport property measurements revealed that the electrical conductivity of the trigonal composites were several times higher compared to the cubic counterparts while thermal conductivity is lower but in the range of previously reported Mg-Si-Sn materials. The Seebeck coefficients of the trigonal composites is relatively low leading to low *ZT* values. The *ZT* of the cubic samples however is higher than previously reported for nominally un-doped Mg-Si-Sn solid solutions which is promising for thermoelectric applications.

## Acknowledgements

UC Lab Fees Program (grant ID#12-LR-239003) is gratefully acknowledged for financial support. Sample synthesis and processing was accomplished with funding from "Spins and Heat in Nanoscale Electronic Systems (SHINES)", an Energy Frontier Research Center funded by the U.S. Department of Energy, Office of Science, Basic Energy Sciences (BES) under award # SC0012670.

**Table 1:**

| Sample ID | Planetary ball milled | CAPAD processing temperature (°C) | Bulk density (g/cm$^3$) | Mg-Si-Sn Phase present |
|---|---|---|---|---|
| PBM 0 | Yes | 355 | 3.68 | Trigonal + Cubic |
| PBM 1 | Yes | 415 | 3.82 | Trigonal + Cubic |
| PBM 2 | Yes | 460 | 3.92 | Trigonal + trace amount of cubic |
| PBM 3 | Yes | 500 | 3.81 | Trigonal |
|  |  |  |  |  |
| No PBM I | No | 600 | 3.12 | Cubic |
| No PBM II | No | 650 | 3.24 | Cubic |
| No PBM III | No | 700 | 3.25 | Cubic |

# Figures

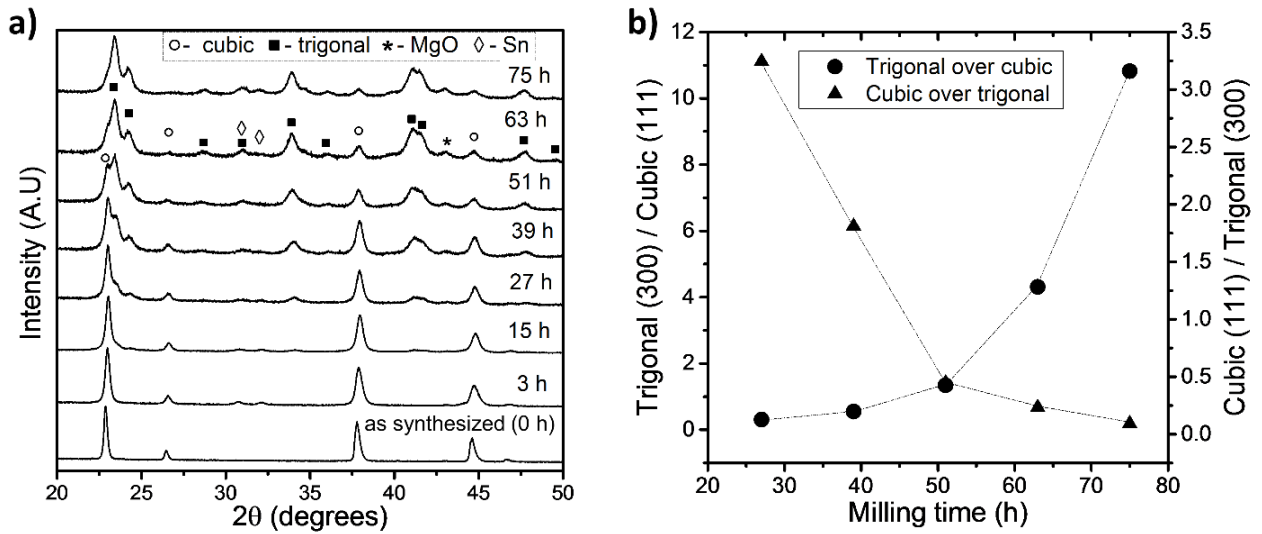

**Figure 1:** Influence of the high-energy planetary ball milling (PBM) on the phase transformation in pre formed Mg-Si-Sn system. (a) XRD powder patterns of the planetary ball milled powders of the $Mg_2Si_{0.1}Sn_{0.9}$ composition with increasing milling hours; (b) integral intensity ratios of trigonal (300) and Cubic (111) vs milling duration. The dotted lines are guide for the eyes.

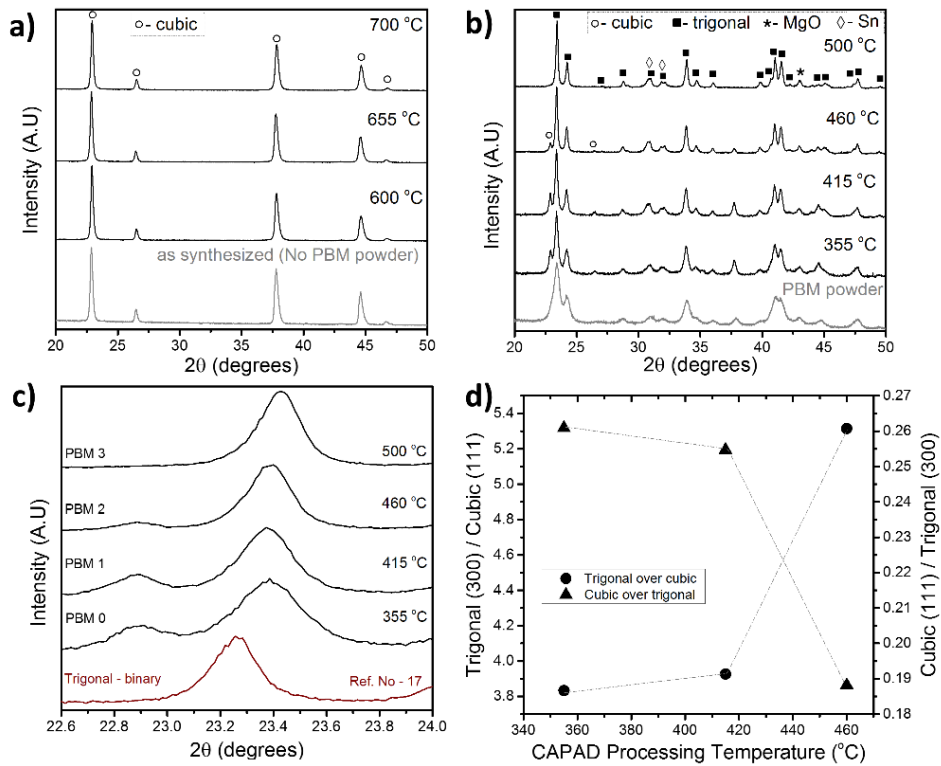

**Figure 2:** Role of PBM and CAPAD processing on completion of phase transformation. (a) XRD of the as-synthesized $Mg_2Si_{0.1}Sn_{0.9}$ composition powders and CAPAD processed specimens at varying temperatures with a hold time of 5 minutes under 112 MPa; (b) XRD patterns of the $Mg_2Si_{0.1}Sn_{0.9}$ composition PBM powder after CAPAD processed; (c) magnified

region corresponds to the trigonal (300) of Mg-Si-Sn system with trigonal (300) of Mg-Sn system; (d) integral intensity ratios of trigonal (300) and cubic (111) vs CAPAD processing temperature. The dotted lines are guide for the eyes.

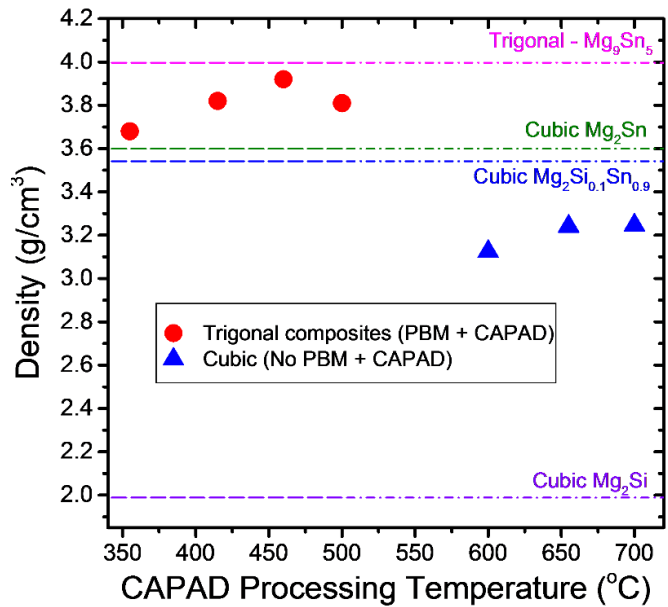

**Figure 3:** Measured bulk density values of CAPAD processed samples with varying CAPAD processing temperature under 112 MPa and a hold time of 5 minutes. Cell data value for cubic $Mg_2Sn$ / $Mg_2Si$ and trigonal $Mg_9Sn_5$ systems are from ICSD references. Theoretical density values of the cubic $Mg_2Si_{0.1}Sn_{0.9}$ (3.542 g/cm$^3$) calculated using the XRD pattern of 600 °C/6h powder.

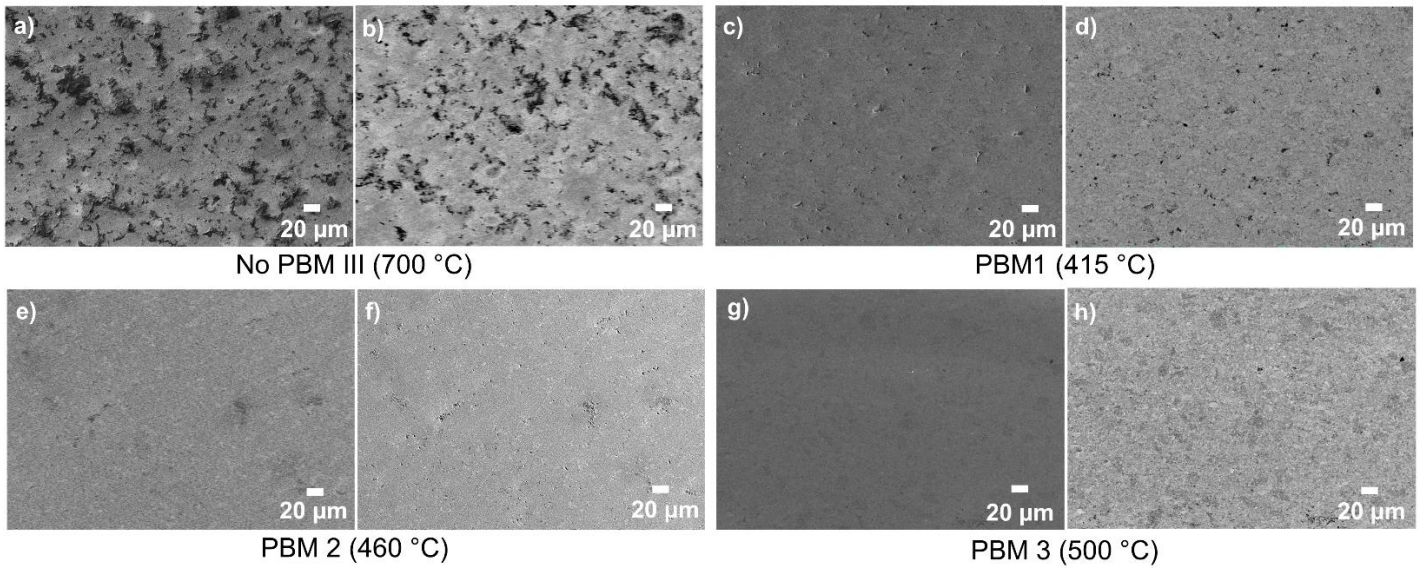

**Figure 4:** Microstructure analysis of the $Mg_2Si_{0.1}Sn_{0.9}$ composition.(a) No PBM III (700 °C) (b) PBM 1 (415 °C), (c) PBM 2 (460 °C), and (d) PBM 3 (500 °C). Secondary electron (SE) images are provided at the first followed by the back scattered images (BSE). CAPAD processing conditions are in the closed bracket.

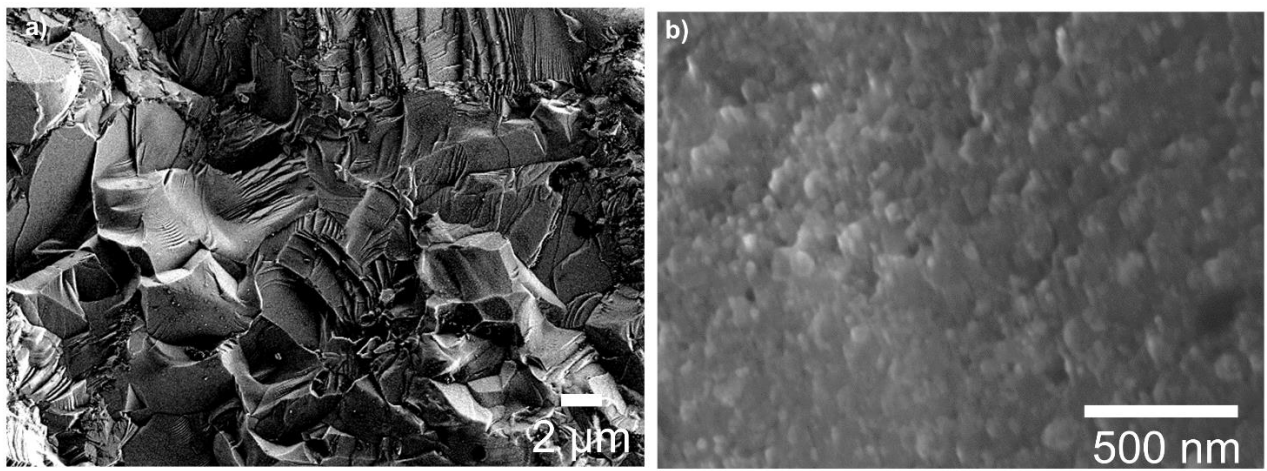

**Figure 5:** Fracture surface of (a) Cubic - No PBM III (CAPAD processed at 700 °C); and (b) Trigonal - PBM 3 (CAPAD processed at 500 °C)

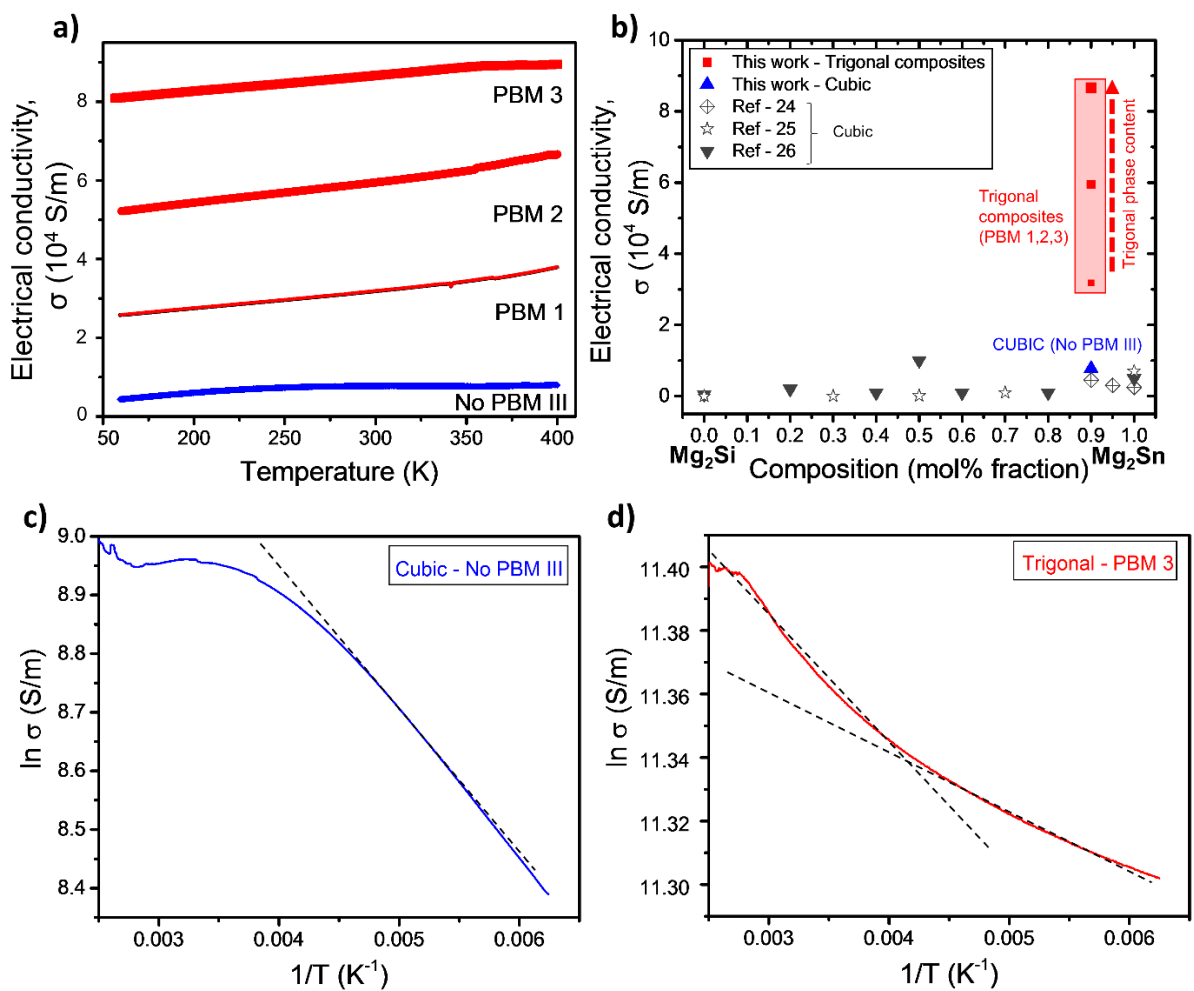

**Figure 6:** Measured Electrical conductivity values of Mg-Si-Sn samples. (a) Selected cubic and trigonal composites with increasing temperature; (b) Electrical conductivity values of the present study at room temperature in comparison with the previously published cubic samples belongs to $Mg_2Si_{(1-X)}Sn_X$ system; (c) the electronic conductivity (ln σ) behavior temperature (1/T) of cubic (No PBM III); and (d) trigonal (PBM 3) samples. The dashed lines show linear dependence ranges used for calculating activation energies, *E*.

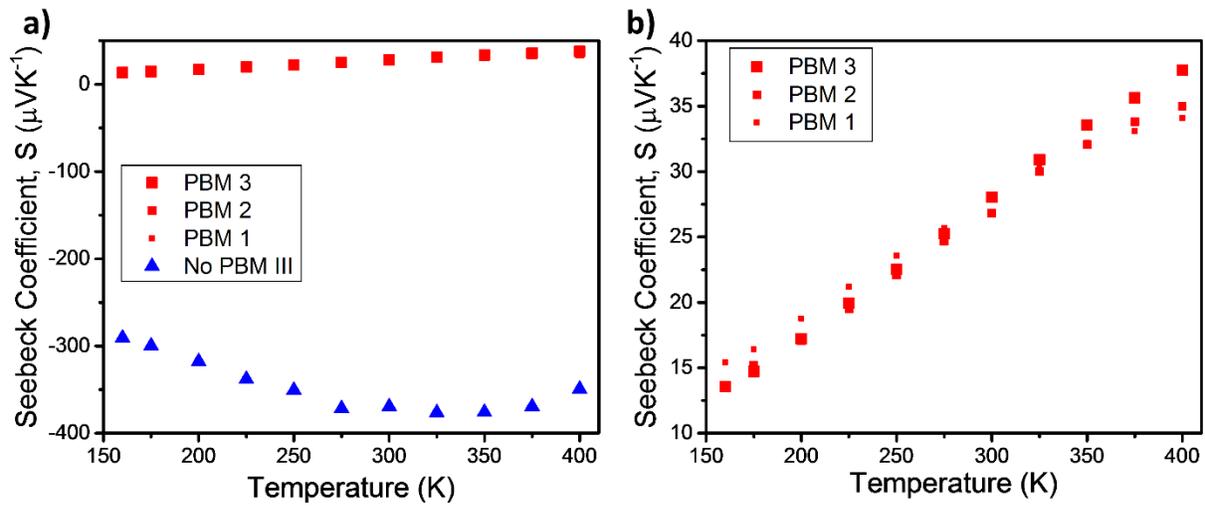

**Figure 7:** Seebeck coefficient values of the (a) trigonal composites, and (b) cubic and trigonal composites measured at various temperatures.

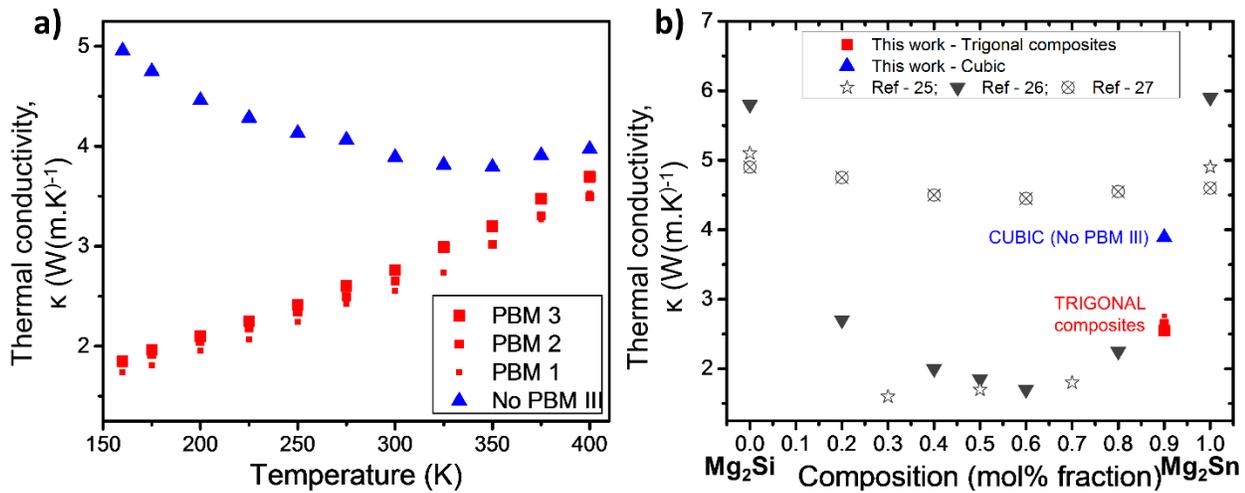

**Figure 8:** Thermal conductivity values of the $Mg_2Si_{0.1}Sn_{0.9}$ composition. (a) Selected cubic and trigonal composites measured at various temperatures. (b) Thermal conductivity values of the present study at room temperature in comparison with the previously published cubic samples belongs to $Mg_2Si_{(1-x)}Sn_x$ system.

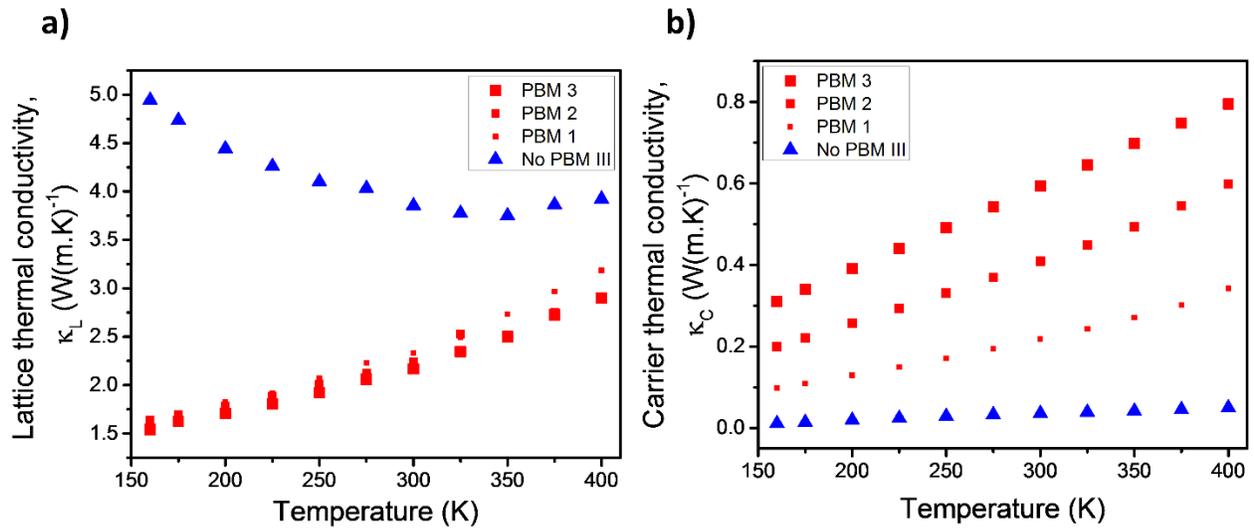

**Figure 9:** (a) Lattice and (b) carrier thermal conductivity values of the present samples at various temperatures.

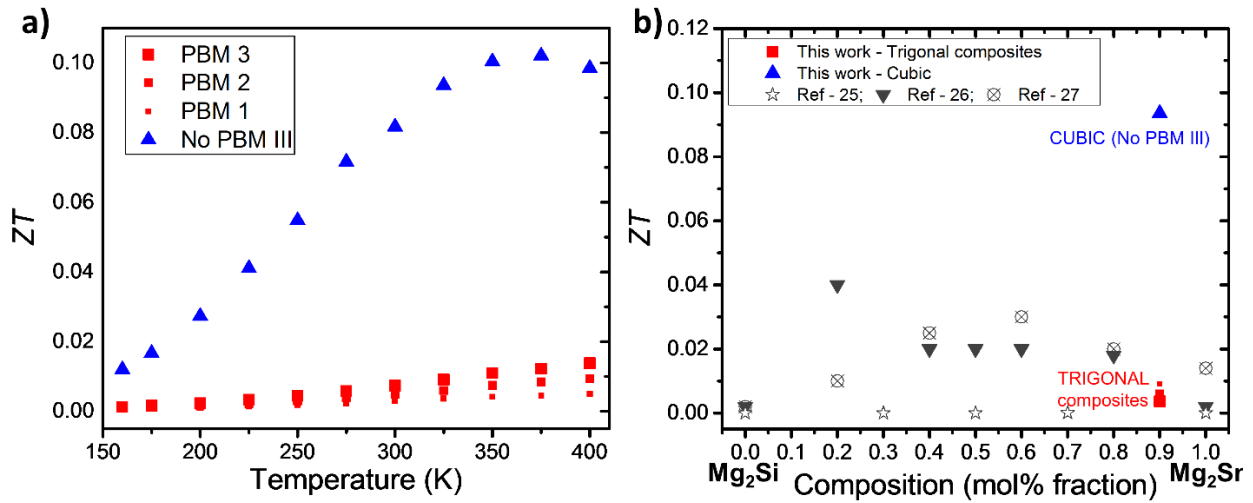

**Figure 10:** (a) Figure of merit, *ZT*, values of the present samples with increasing temperature and (b) *ZT* values of the present study at room temperature in comparison with the previously published cubic samples belongs to $Mg_2Si_{(1-X)}Sn_X$ system.